\documentclass[prl,aps,twocolumn,superscriptaddress,nofootinbib]{revtex4}

\usepackage{epsfig}
\usepackage{amsfonts}
\usepackage{latexsym}
\usepackage{amsmath}
\usepackage{amssymb}
\usepackage{slashed}
\usepackage{longtable}

\usepackage{graphicx}
\usepackage{hyperref}
\usepackage{pstool}

\numberwithin{equation}{section}

 \def\p{\partial}

\newcommand{\bea}{\begin{eqnarray}}
\newcommand{\eea}{\end{eqnarray}}
\newcommand{\be}{\begin{equation}}
\newcommand{\ee}{\end{equation}}
\newcommand{\ba}{\begin{align}}
\newcommand{\ea}{\end{align}}

\newcommand{\prn}[1]{\left ( #1 \right )}

\def\Or[#1]{{\text{O}}\left({#1}\right)}
\def\dotl[#1,#2]{\left\langle #1, #2 \right\rangle}
\def\dotlb[#1,#2]{[ #1, #2 ]}
\def\dotp[#1,#2]{(#1) \cdot (#2)}
\def\aff[#1,#2]{\hat{#1}(#2)}
\def\n4sym{{\cal N}=4 SYM}
\def\>{\rangle}
\def\<{\langle}
\def\weight[#1,#2,#3]{\{(#1),#2,#3\}}
\def\ads[#1]{$\text{AdS}_{#1}$}

  \makeatletter
  \let\over=\@@over \let\overwithdelims=\@@overwithdelims
  \let\atop=\@@atop \let\atopwithdelims=\@@atopwithdelims
  \let\above=\@@above \let\abovewithdelims=\@@abovewithdelims

\usepackage{xcolor}

\begin{document}

\preprint{}

\title{Information Geometry of Hydrodynamics with Global Anomalies}

\author{Piotr Sur\'owka} \affiliation{Center for the Fundamental Laws of Nature, Harvard University, Cambridge, MA 02138, USA}

\begin{abstract}
We construct information geometry for hydrodynamics with global gauge and gravitational anomalies in $1+1$ and $3+1$ dimensions. We introduce the metric on a parameter space and show that turning on non-zero rotations leads to a curvature on the statistical manifold. We calculate the curvature invariant and analyze its divergences, which occur at the transition points of the system. The transition points are universal and expressed in terms of ratios of anomaly coefficients.
\end{abstract}

\maketitle
\emph{Introduction.}---Relativistic fluids constitute the quark-gluon plasma (QGP), which was present during the early stages of the Universe and now is recreated in the heavy-ion experiments at Relativistic Heavy Ion Collider (RHIC) and Large Hadron Collider (LHC). It was realized that transport properties of fluids might be modified if the underlying constituents break classical global symmetries leading to the quantum anomalies \cite{Erdmenger:2008rm,Banerjee:2008th,Son:2009tf,Landsteiner:2011cp}. As typical liquids are strongly coupled the number of methods to study their properties is limited. However, the topological nature of anomalies makes transport due to anomalies very special. In particular it is non-dissipative and insensitive to the changes of the coupling, which makes the analysis analytically tractable. In fact we can not only use the weak coupling techniques such as semiclassical kinetic theory \cite{Son:2012wh,Stephanov:2012ki}, but we can write down the full partition function for this system \cite{Jensen:2012jh,Banerjee:2012iz,Loganayagam:2012pz,Loganayagam:2012zg,Ng:2014sqa}.

Partition functions are primary objects in studying field theories and statistical systems. Unfortunately the number of models, for which partition functions are known exactly is very limited and in order to get a non-perturbative answer one usually invokes underlying integrability or supersymmetry of the model. However, most models do not have such a symmetry and the partition function is known only perturbatively in a limited range of the parameter space. Anomalous hydrodynamics is an exception in this respect. The powerful constraints coming from the connection of QFT symmetry breaking and the laws of thermodynamics make the transport non-dissipative and fixed purely in terms of field theory data. As a result the partition function of the theory in the hydrodynamic regime can be constructed analytically. This gives as a closed form expression for the Gibbs probability distribution for the anomalous state. The statistical distributions can be viewed as geometrical manifolds, which has a non-zero curvature and possibly nontrivial phase structure if the system is interacting. This represents a new quantitative tool for the study of fluctuation phenomena, know as information geometry.

The aim of this paper is to use information geometric methods to investigate the properties of the statistical manifold of a system of Weyl fermions in the hydrodynamic regime. In the long wavelength expansion the system has transport properties fixed by anomalies that affect chiral fluid constituents. This can be visualised as a separation of chiral particles given a non-zero rotation, which is known as chiral vortical effect. We will calculate critical points for this system, which should be applicable to strongly coupled chiral fermionic systems such as QGP or Weyl semimetals \cite{Turner:2013kp}. For some values of the parameter the fluids are unstable, which was also noticed in the kinetic theory \cite{Akamatsu:2013pjd}. Similar instabilities were also found in the context of the early universe \cite{Joyce:1997uy,Boyarsky:2011uy}. New connection between QFT anomalies and information geometry as well as the novel perspective on chiral instabilities are the main results of this paper.

\emph{Information Geometry}---The field of information geometry was developed in order to study the phase space of statistical systems using geometry \cite{PhysRevA.20.1608,amari2000methods,RevModPhys.67.605,Balian19861,bengtsson2008geometry,barbaresco}. A given statistical ensemble is represented as a point on a Riemannian manifold. This manifold is endowed with a metric which is precisely the Fisher-Rao information metric. In such a geometrization a scalar curvature $\mathcal{R}$ plays a central role and contains the information about phase transitions. In order to see how it works let us start with a statistical system immersed in a heat bath in thermal equilibrium. The system is characterised by a set of thermodynamic parameters $\beta _i$ which include inverse temperature and generalized chemical potentials for the conserved quantities. One can write down a Gibbs measure for this system
\be
p(x|\beta)= \exp \left( - \sum _{i}  \beta ^i H_i(x) -\ln \mathcal{Z}(\beta) \right),
\ee
where $H_i(x)$ include the hamiltonian and conserved currents, $\mathcal{Z}(\beta)$ is the partition function. Given a Gibbs probability measure we can define Fisher information matrix
\be
G_{ij}(\beta)=-\left< \frac{\p ^2 \ln p(x|\beta)}{\p \beta_i \p \beta _j}  \right>.
\ee
It was suggested by Rao \cite{rao1945} that this is a metric and is now known as the Fisher-Rao information metric. This metric can be proven to be unique \cite{centsov1982statistical}. From the physical perspective the most important role is played by the Ricci scalar curvature $\mathcal{R}$. If the system is non-interacting the curvature is zero and if the system exhibits a phase transition $\mathcal{R}$ blows up at the transition point.

The thermodynamic systems investigated with the use of information geometry include discrete Ising models in a magnetic field \cite{PhysRevA.39.6515,Brody2003207}, ideal Bose and fermi gases \cite{janyszek90} and more recently classical systems of anyons \cite{PhysRevE.78.021127}, which provide an example of a strongly coupled theory. The thermodynamic curvature for ideal Bose and Fermi gases has been shown to have a different sign. We will show that anomalous hydrodynamics is a theory, in which analysis based on information geometry is possible and sheds light on the phase space of interacting systems with global anomalies. The transition points exhibit universal behavior and are given in terms of anomaly coefficients.

\emph{Partition functions of anomalous hydrodynamics}---Hydrodynamics of relativistic fluids can be described by temperature $T$ and velocity field $u^\mu$, together with the normalization condition $u^\mu u_\mu =-1$. Moreover, if we have conserved charges and angular momenta corresponding chemical potentials and angular velocities are hydrodynamic degrees of freedom as well. The equations of motion are given as conservation laws of energy-momentum tensor and the current
\be
D_\mu T^{\mu \nu}=0,
\ee
\be
D_\mu J^\mu=0.
\ee
The energy-momentum tensor and the current may be expressed through the constitutive relations that involve the derivative expansion of hydrodynamic coefficients and their gradients. This expansion is subject to several constraints. First of all it must be in accordance with the symmetries of the underlying microscopic theory. Moreover, the second law of thermodynamics together with the Onsager relations leads to restrictions on the allowed tensor structures \cite{Jensen:2012jh,Banerjee:2012iz}. In the non-dissipative systems the stress-tensor and currents can be generated from a single generating functional. To make a direct connection between fluids with anomalies and information geometry in $1+1$ and $3+1$ dimension it is convenient to take fluid configurations on $R\times S^1$  and $R\times S^3$ \cite{Loganayagam:2012zg}. Before writing down the expressions for the partition functions we introduce the notion of anomaly polynomials $\mathcal{P}_{anom}(F,R)$. They are functions of gauge field strength and curvature, which represent a compact way to describe anomalies. It was argued in \cite{Loganayagam:2012pz,Jensen:2012kj,Jensen:2013kka} that the anomaly induced transport can be captured by a closely related polynomial object ${\mathfrak{F}}^\omega_{anom}(\mu,T)$, provided we do the the following substitution
\be
{\mathfrak{F}}^\omega_{anom} = \mathcal{P}_{anom}\prn{F\mapsto \mu\ ,\ p_{_1}(R)\mapsto -\beta^{-2} ,\ p_{k>1}(R)\mapsto 0},
\ee
where $p_{_1}(R)$ is the first Pontryagin class of space-time curvature. Subsequently the polynomial object ${\mathfrak{F}}^\omega_{anom}$ was connected to the helicity of the thermal state. We are now interested in calculating the partition function of a fluid rotating on a $2n$-dimensional sphere
\be
\mathcal{Z}=\mathrm{Tr} \exp \left[ -\beta (H-\mu N-\sum _{a=1}^n \Omega_a L_a )\right],
\ee
where $N$ is the particle number, $\Omega _a$ denote angular velocities for mutually commuting angular momenta $L_a$. For a general theory evaluating the partition function would be an impossible task, however, in the case of anomalous fluids this can be done analytically. We consider the limit in which the radius of the sphere $R$ is large and the angular velocities are small. The effective action for a conformal fluid takes the form
\bea
\label{eq:partf}
\mathcal{W}\equiv\ln \mathcal{Z}&=&
\beta \left[ p \frac{\mathrm{Vol}_{S^{2n-1}}}{\prod _{a=1}^n(1-R^2\Omega_a^2)}+\ldots \right]\nonumber\\
&&-\beta \left[
{\mathfrak{F}}^\omega_{anom}\prod _{a=1}^n\left(\frac{2\pi R^2 \Omega _a}{1-R^2\Omega_a^2}\right) +\ldots \right],
\eea
where $p$ is the fluid pressure. The first term corresponds to the leading, non-universal, parity-even hydrodynamics and the second captures the leading parity-odd hydrodynamics. The exact equation of state that allows one to express pressure as a function of temperature and chemical potential is known only for a limited number of examples that include $1+1$-dimensional fluids in the Cardy regime or fluids with gravity duals that allow non-perturbative calculations. However, here we focus on the parity-odd part, which is universal and depends only on the anomalies of the theory.

\emph{Information Geometry of $1+1$-dimensional fluids}---Our first example, which we analyze using methods taken from the information geometry will be fluids on $R\times S^1$. The anomaly polynomial in two dimensions allows one to determine ${\mathfrak{F}}^\omega_{anom}$, which is given by
\be
{\mathfrak{F}}^\omega_{anom}=c_A  \mu^2 +c_g  \beta ^{-2} ,
\ee
where the
\be
c_A =-\frac{1}{2! 2\pi}\sum \chi_i q_i,
\ee
\be
c_g=-\frac{2\pi}{4! }\sum \chi_i q_i
\ee
represent gauge and gravitational anomaly coefficients in two dimensions and $\chi_i$ is the chirality of the fermionic species. The partition function \eqref{eq:partf} reads
\be
\label{eq:partf2d}
\mathcal{W}=
\beta \left[ p \frac{2\pi R}{(1-R^2\Omega ^2)}-{\mathfrak{F}}^\omega_{anom}\frac{2\pi R^2 \Omega }{1-R^2\Omega ^2} \right].
\ee
We see that the partition function has a non-universal term that depends on the details of our theory through $p$ and the universal term that is proportional to the chiral vortical coefficient ${\mathfrak{F}}^\omega_{anom}$. We are interested in the universal parity-odd piece
\be \label{eq:partf2danom}
\mathcal{W}_{\mathrm{anom}}=-\beta {\mathfrak{F}}^\omega_{anom}\frac{2\pi R^2 \Omega}{1-R^2\Omega ^2} .
\ee
Our goal is to analyze the properties of a statistical manifold $\mathcal{M}$, which is in general parameterized by a set of intensive parameters $\{\beta ,\mu,\Omega\}$. We also consider lower dimensional isosurfaces of that manifold. If $\mathcal{M}$ is one-dimensional the curvature is always zero. Therefore in order to get a non-trivial structure we study at least two-dimensional submanifolds $\mathcal{M} ^{\beta,\mu}$, $\mathcal{M} ^{\beta, \Omega} $, $\mathcal{M}^{\mu, \Omega} $ or three-dimensional $\mathcal{M}^{\beta, \mu, \Omega} $. We introduce the anomalous Fisher-Rao information metric
\be \label{eq:anommetric}
G_{ij}^{\mathrm{anom}}=\frac{\p ^2 \mathcal{W}_{\mathrm{anom}}}{\p \beta_i \p \beta _j},
\ee
which encodes the geometric properties of statistical manifolds related to anomalies. We evaluate the leading-order metric \eqref{eq:anommetric} in $1/R$ expansion using $1+1$-dimensional partition function \eqref{eq:partf2danom}, for which define the Ricci curvature scalar $\mathcal{R}_{anom}$. It is divergent at the critical point. We list the curvature invariants and the corresponding critical points in Table \ref{tbl:FI2d}. The critical temperature is always proportional to the ratio of anomaly coefficients.
We note that the utility of putting the system on a sphere comes from the fact that it provides an infrared regulator. A rigid rotation in flat space-time breaks down at large radius away from the centre since the velocities involved eventually exceed the relativistic limit. Moreover, the partition function becomes divergent. To see this we set
\be
\Omega = \frac{\mathbf{P}}{R},
\ee
where $\mathbf{P}$ is the flat-space momentum. As a result $\mathcal{W}_{\mathrm{anom}}$ is proportional to $R$. The Ricci scalar in the parameter space vanishes, which confirms the standard lore that a chiral fluid on an infinite line undergoes no phase transition. We note, however, that if we fix momentum and consider a finite volume, the flat-space Ricci scalar is non-zero and differs from the leading-order sphere Ricci scalar only by a factor in front, which will not affect the transition point. Therefore our result for the transition points remains valid at finite volume in flat space, keeping in mind that the dimension of the statistical manifold is lower. This a consequence of the reduction of the symmetry algebra in the decompactification limit \cite{Loganayagam:2012zg}. In $1+1$ dimensions the flat-space statistical manifold is $\mathcal{M} ^{\beta,\mu}$.
\paragraph{}
\mbox{}%
\begin{longtable*}{|l|c|c|}
\hline
Manifold $\mathcal{M}$ & Ricci scalar $\mathcal{R}_{\mathrm{anom}}$ & Critical point \\
\hline
$\mathcal{M} ^{\beta,\mu}$ & $-\frac{6 \pi ^2 \beta  c_g \Omega }{\left(\pi ^2 c_g-3 \beta ^2 c_A \mu ^2\right)^2}$  & $T_c= \frac{\sqrt{3}}{\pi} \left(\frac{ c_A}{c_g}\right)^{1/2}\mu$ \\
$\mathcal{M} ^{\beta, \Omega} $ & $\frac{24 \beta ^4 c_A  \Omega  \left(3 \beta ^5 c_A ^2 \mu ^4-\pi ^2 \beta ^2 c_A  c_g \mu ^3+\pi ^4 c_g^2 (\beta
   -\mu )\right)}{\left(-3 \beta ^4 c_A ^2 \mu ^4+6 \pi ^2 \beta ^2 c_A  c_g \mu ^2+\pi ^4 c_g^2\right)^2}$ & $T_c= \left(\frac{2\sqrt{3}-3}{3}\right)^{1/2} \left(\frac{ c_A}{c_g}\right)^{1/2} \mu $\\
$\mathcal{M}^{\mu, \Omega} $ & $\frac{3 \beta  \Omega }{\pi ^2 c_g}$ & no critical point\\
$\mathcal{M}^{\beta,\mu, \Omega} $ & $\frac{\beta  \Omega  \left(\beta ^6 \left(-c_A^3\right) \mu ^6-25 \beta ^4 c_A^2 c_g  \mu ^4+33 \beta ^2 c_A c_g ^2 \mu ^2+9 c_g ^3\right)}{2 \pi
   \left(\beta ^4 c_A^2 \mu ^4+6 \beta ^2 c_A c_g  \mu ^2-3 c_g ^2\right)^2}$ & $T_c= \left(\frac{3+2\sqrt{3}}{3}\right)^{1/2} \left(\frac{ c_A}{c_g}\right)^{1/2} \mu $\\
\hline
\caption{Scalar curvature invariants and corresponding critical points for anomalous hydrodynamics in $1+1$ dimensions}\label{tbl:FI2d}
\end{longtable*}

\emph{Information Geometry of $3+1$-dimensional fluids}---We now turn to more phenomenologically interesting case of chiral fluids in $3+1$ dimensions. Fluids with chirality imbalance are expected to appear in quark-gluon
plasma, which is a phase of extremely hot matter consisting of quarks and gluons. This chirally imbalanced matter is characterized by different densities of quarks with opposite helicities. It is now widely appreciated that during the expansion after the collision the plasma reaches thermodynamic equilibrium and can be described by hydrodynamics. We therefore expect that various effects related to the anomaly-induced transport will be present. So far most studies have been devoted to understanding the theoretical aspects of that transport. In this paper we want to ask how it affects the phase structure. The first principles studies of QCD phases are limited by our ignorance how to analyze strongly coupled systems. At asymptotically large energies, perturbative QCD can be used. First-principle QCD calculations in lower energies can only be done on the lattice and are restricted to vanishing (vector) chemical potentials. We have argued that the information geometry can be useful to study transition points of anomalous hydrodynamics. This is possible because the transport is non-dissipative and in order to determine the partition function we only need to invoke free field theory methods. We are interested in the anomalous part of the partition function and we do not need to assume any particular equation of state specific to the underpinning microscopic theory. Therefore we can extract universal properties of the phase transitions for anomalous fluids that depend only on the anomaly coefficients.

In $3+1$ dimensions the anomaly structure is different than in $1+1$. In particular there is no pure gravitational anomaly but instead the so-called mixed anomaly is present, which is reflected in the anomaly polynomial and
\be
{\mathfrak{F}}^\omega_{anom}= \tilde{c}_A  \mu^3 +c_m  \frac{\mu}{\beta ^{2}} ,
\ee
with
\be
\tilde{c}_A =-\frac{1}{3! (2\pi)^2}\sum \chi_i q_i^3,
\ee
\be
c_m=-\frac{1}{4! }\sum \chi_i q_i
\ee
being gauge and mixed anomaly coefficients in $3+1$ dimensions. We can now write down the the partition function using \eqref{eq:partf}
\bea
\mathcal{W}&=&
\beta \left[ p \frac{2\pi ^2 R^3}{(1-R^2\Omega_1 ^2)(1-R^2\Omega_2 ^2)}\right] \nonumber\\
&&-\beta \left[{\mathfrak{F}}^\omega_{anom}\frac{2\pi R^2 \Omega_1}{1-R^2\Omega_1 ^2} \frac{2\pi R^2 \Omega_2}{1-R^2\Omega_2 ^2}\right],
\eea
where the second line corresponds to chiral fluid. In $3+1$ dimensions on spheres we have 2 Cartan generators for the rotation group, therefore we have one more coordinate in our statistical manifold. Note, however, that the partition function is symmetric under exchanging the angular velocities. Keeping that in mind we calculate the leading-order anomalous Fisher-Rao metric \eqref{eq:anommetric} for a large $R$ and list associated Ricci scalars together with the corresponding critical points in Tables \ref{tbl:R4d} and \ref{tbl:CP4d}. These critical points exhibit universal behavior as they depend only on the combinations of anomaly coefficients. In order to take the flat space-time limit we need to set $\Omega _2= \mathbf{P}/R$, which manifests the fact that the symmetry algebra is reduced in the flat-space limit. Similarly to the $1+1$-dimensional case, for a fixed momentum the Fisher-Rao metric and the finite volume Ricci scalar differ only by a factor that will not affect the transition point. As a result the transition points of statistical manifolds that do not depend on $\Omega _2$ will remain unchanged in flat space-time. We find critical points for most of sub-maniflods including $\mathcal{M}^{\beta, \mu }$ and $\mathcal{M}^{\beta, \mu, \Omega _1}$, which should be the most accessible in the heavy-ion collisions. It would be interesting to investigate the critical behavior using other methods such as lattice simulations and, most importantly, confirm it experimentally. We leave the discussion of possible phenomenological observables for future research.
\paragraph{}
\mbox{}%
\begin{longtable*}{|l|c|}
\hline
Manifold $\mathcal{M}$ & Ricci scalar $\mathcal{R}_{\mathrm{anom}}$  \\
\hline
$\mathcal{M} ^{\beta,\mu}$ & $-\frac{6 \beta ^3 \tilde{c}_A  c_m \mu   \left(3 \beta ^2 \tilde{c}_A  \mu ^2+c_m\right) \Omega _1 \Omega _2}{\pi ^2 \left(9 \beta ^4 \tilde{c}_A ^2 \mu ^4-18
   \beta ^2 \tilde{c}_A  c_m \mu ^2+c_m^2\right)^2} $   \\
$\mathcal{M} ^{\beta, \Omega_i} $ & $\frac{\beta ^3 \tilde{c}_A c_m \mu   \left(\beta ^2 \tilde{c}_A \mu ^2+3 c_m\right) \Omega _i \Omega _j}{\pi ^2 \left(-\beta ^4 \tilde{c}_A^2 \mu
   ^4+6 \beta ^2 \tilde{c}_A c_m \mu ^2+3 c_m^2\right)^2}, \qquad \Omega_i \neq \Omega_j$  \\
$\mathcal{M}^{\mu, \Omega _i} $ & $\frac{3 \beta ^3 \tilde{c}_A c_m \mu   \left(\beta ^2 \tilde{c}_A \mu ^2-c_m\right)\Omega _i \Omega _j}{\pi ^2 \left(-3 \beta ^4 \tilde{c}_A^2 \mu ^4-6 \beta
   ^2 \tilde{c}_A c_m \mu ^2+c_m^2\right)^2}, \qquad \Omega_i \neq \Omega_j$  \\
$\mathcal{M}^{ \Omega _1,\Omega _2} $ & $0$  \\
$\mathcal{M}^{\beta, \mu, \Omega _i} $ & $\frac{\beta  \left(27 \beta ^{10} \tilde{c}_A^5 \mu ^{10}+21 \beta ^8 \tilde{c}_A^4 c_m \mu ^8-354 \beta ^6 \tilde{c}_A^3 c_m^2 \mu ^6+90
   \beta ^4 \tilde{c}_A^2 c_m^3 \mu ^4-41 \beta ^2 \tilde{c}_A c_m^4 \mu ^2+c_m^5\right)\Omega _i \Omega _j}{16 \pi ^2 \mu  \left(c_m-\beta ^2 \tilde{c}_A \mu ^2\right)^2 \left(-3
   \beta ^4 \tilde{c}_A^2 \mu ^4-6 \beta ^2 \tilde{c}_A c_m \mu ^2+c_m^2\right)^2}, \qquad \Omega_i \neq \Omega_j$  \\
$\mathcal{M}^{\beta, \Omega _1, \Omega _2} $ & $\frac{\beta  \left(\beta ^8 \text{c1}^4 \mu ^8+\beta ^6 \text{c1}^3 \tilde{c}_A \mu ^6+73 \beta ^4 \text{c1}^2 \tilde{c}_A^2 \mu ^4+63 \beta ^2
   \text{c1} \tilde{c}_A^3 \mu ^2+6 \tilde{c}_A^4\right)\Omega _1 \Omega _2}{16 \pi ^2 \mu  \left(\beta ^2 \tilde{c}_A \mu ^2+\tilde{c}_A\right) \left(\beta ^4 \left(-\tilde{c}_A^2\right) \mu ^4+5 \beta ^2
   \tilde{c}_A \tilde{c}_A \mu ^2+2 \tilde{c}_A^2\right)^2}$  \\
$\mathcal{M}^{ \mu,\Omega_1, \Omega_2} $ & $\frac{\beta   \left(45 \beta ^6 \tilde{c}_A^3 \mu ^6+15 \beta ^4 \tilde{c}_A^2 \tilde{c}_A \mu ^4-45 \beta ^2 \tilde{c}_A \tilde{c}_A^2 \mu
   ^2+\tilde{c}_A^3\right)\Omega _1 \Omega _2}{16 \pi ^2 \tilde{c}_A \mu  \left(\tilde{c}_A-3 \beta ^2 \tilde{c}_A \mu ^2\right)^2 \left(\beta ^2 \tilde{c}_A \mu ^2+\tilde{c}_A\right)}$  \\
$\mathcal{M}^{\beta, \mu, \Omega_1, \Omega_2} $ & $\frac{3 \beta \left(9 \beta ^{12} \tilde{c}_A^6 \mu ^{12}+192 \beta ^{10} \tilde{c}_A^5 \tilde{c}_A \mu ^{10}+57 \beta ^8 \tilde{c}_A^4 \tilde{c}_A^2 \mu
   ^8-168 \beta ^6 \tilde{c}_A^3 \tilde{c}_A^3 \mu ^6+187 \beta ^4 \tilde{c}_A^2 \tilde{c}_A^4 \mu ^4-24 \beta ^2 \tilde{c}_A \tilde{c}_A^5 \mu ^2+3 \tilde{c}_A^6\right)\Omega _1 \Omega _2}{4 \pi ^2 \mu
   \left(\beta ^2 \tilde{c}_A \mu ^2+\tilde{c}_A\right) \left(3 \beta ^6 \tilde{c}_A^3 \mu ^6+21 \beta ^4 \tilde{c}_A^2 \tilde{c}_A \mu ^4-11 \beta ^2 \tilde{c}_A \tilde{c}_A^2 \mu ^2+3
   \tilde{c}_A^3\right)^2}$   \\
\hline
\caption{Scalar curvature invariants of anomalous statistical manifolds in $3+1$ dimensions}\label{tbl:R4d}
\end{longtable*}

\paragraph{}
\mbox{}%
\begin{longtable*}{|l|c|}
\hline
Manifold $\mathcal{M}$ & Critical point \\
\hline
$\mathcal{M} ^{\beta,\mu}$ & $T_c=\left(\frac{3(3-2\sqrt{2})\tilde{c}_A}{c_m}\right)^{1/2}\mu$ ;$\qquad  T_c=\left(\frac{3(3+2\sqrt{2})\tilde{c}_A}{c_m}\right)^{1/2}\mu$  \\
$\mathcal{M} ^{\beta, \Omega_i} $ & $ T_c=\left(\frac{(2\sqrt{3}-3)\tilde{c}_A}{3 c_m}\right)^{1/2}\mu$ \\
$\mathcal{M}^{\mu, \Omega _i} $ & $ \mu_c=\left(\frac{(2\sqrt{3}-3)c_m}{3 \tilde{c}_A}\right)^{1/2}T$ \\
$\mathcal{M}^{ \Omega _1,\Omega _2} $ & no critical point \\
$\mathcal{M}^{\beta, \mu, \Omega _i} $ &$T_c=\left(\frac{\tilde{c}_A}{c_m}\right)^{1/2}\mu$;$\qquad T_c=\left(\frac{(3+2\sqrt{3})\tilde{c}_A}{3c_m}\right)^{1/2}\mu$  \\
$\mathcal{M}^{\beta, \Omega _1, \Omega _2} $ & $T_c=\left(\frac{(\sqrt{33}-5)\tilde{c}_A}{4 c_m}\right)^{1/2}\mu$   \\
$\mathcal{M}^{ \mu,\Omega_1, \Omega_2} $ & $\mu_c=\left(\frac{c_m}{3 \tilde{c}_A}\right)^{1/2}T$ \\
$\mathcal{M}^{\beta, \mu, \Omega_1, \Omega_2} $ & no critical point
 \\
\hline
\caption{Critical points of anomalous hydrodynamics in $3+1$ dimensions}\label{tbl:CP4d}
\end{longtable*}

\emph{Conclusions}---In this article, we constructed information geometry for hydrodynamics with global gauge and gravitational anomalies, which is a strongly coupled system. This connects two disciplines and we believe that our formulation will cross-fertilize both of them. From the information geometry point of view we hope to use the robustness od anomaly related phenomena to learn about properties of statistical manifolds. This may lead to a deeper understanding of quantum theories in terms of information geometry, where anomalies serve as a benchmark in analyzing the probabilistic properties of field theories \cite{Balasubramanian:2014bfa} and the process of constructing probability distributions given geometric data \cite{Clingman:2015lxa}. It may also help to see how anomalies interplay with entanglement entropy and identify possible universal contributions to it.

From the point of view of hydrodynamics the geometric formulation offers tools to study phase space of the evolving anomalous system. We restricted our analysis to the case of chiral vortical effect. A new structure may appear if we introduce external magnetic field, which will trigger the chiral magnetic effect \cite{Fukushima:2008xe}. In this case the statistical manifold will depend on the magnetic field. It might induce new phase transition. In addition to that information geometry can give new insight into the dynamics of fluids described by point vortex models, which can be formulated in statistical terms \cite{majda2006non-linear}. Physical applications of such models include superfluids \cite{pismen1999vortices,Lucas:2014tka} cold atoms \cite{PhysRevLett.83.2498}, and fractional quantum hall effect \cite{Wiegmann:2013hca}.

Finally, since strongly coupled fluids can be mapped to black holes via AdS/CFT correspondence, information geometric methods along the lines suggested in \cite{Aman:2003ug} could be used to confirm the critical behavior in the dual formulation \cite{Erdmenger:2008rm,Banerjee:2008th}.

 The author thanks F.~Barbaresco, J.~Erdmenger, J.~Heckman, K.~Jensen, D.~Kharzeev, and J.~Shock, for useful discussions and comments and the Max Planck Institute for Physics for the hospitality and partial support during the completion of this work. this work was supported by a Marie Curie International Outgoing Fellowship, grant number PIOF-GA-2011-300528.
\bibliographystyle{apsrev4-1}

\bibliography{anomfishinf-bib}

\end{document}